\documentclass[a4paper]{article}

\usepackage{INTERSPEECH2022}

\usepackage{array}
\usepackage{tabularx}
\usepackage{tabulary}
\usepackage{makecell}

\usepackage{multirow}
\usepackage{booktabs}
\usepackage[skip=1pt]{caption}
\usepackage{subcaption}
\usepackage{threeparttable}

\usepackage[inline, shortlabels]{enumitem}
\usepackage[dvipsnames]{xcolor}

\usepackage{textcomp}
\usepackage[normalem]{ulem}

\usepackage[natbib, bibencoding=utf8, citestyle=numeric, bibstyle=ieee, maxbibnames=999, maxcitenames=2, mincitenames=1, sortcites]{biblatex}
\bibliography{bibliography}

\newcommand{\ie}{i.\,e., }
\newcommand{\wrt}{w.\,r.\,t.\ }

\newcommand{\wav}{{w2v2-xlsr}}

\newcommand{\fscore}{{F\textsubscript{1} score\ }}

\usepackage[acronym, shortcuts, nohypertypes={acronym}]{glossaries}
\newacronym{AI}{AI}{artificial intelligence}
\newacronym{ASR}{ASR}{automatic speech recognition}
\newacronym{compare}{{ComParE}}{Interspeech Computational Paralinguistics ChallengE feature set}
\newacronym{COPD}{COPD}{chronic obstructive pulmonary disease}
\newacronym{egemaps}{{eGeMAPS}}{extended Geneva minimalistic acoustic parameter set}
\newacronym{MAUS}{MAUS}{Munich AUtomatic Segmentation}
\newacronym{ML}{ML}{machine learning}
\newacronym{NuS}{NuS}{the North Wind and the Sun [\textit{Der Nordwind und die Sonne}]}
\newacronym{RBF}{RBF}{radial basis function}
\newacronym{SHAP}{SHAP}{SHapley Additive exPlanations}
\newacronym{SVM}{SVM}{support vector machine}
\newacronym{UAR}{UAR}{unweighted average recall}

\usepackage[capitalise, nameinlink, noabbrev]{cleveref}

\title{
    Distinguishing between pre- and post-treatment in the speech of\\patients with chronic obstructive pulmonary disease
}

\name{
    Andreas Triantafyllopoulos$^1$, 
    Markus Fendler$^3$,
    Anton Batliner$^1$,
    Maurice Gerczuk$^1$,\\
    Shahin Amiriparian$^1$,
    Thomas M.\ Berghaus$^{3,4}$,
    and Bj\"orn W.\ Schuller$^{1,2}$
}
\address{
  $^1$Chair of Embedded Intelligence for Health Care and Wellbeing, University of Augsburg, Germany\\
  $^2$GLAM -- Group on Language, Audio, \& Music, Imperial College, UK\\
  $^3$Department of Cardiology, Respiratory Medicine and Intensive Care, University Hospital Augsburg, University of Augsburg, Germany    \\
  $^4$Ludwig-Maximilians-University Munich, Germany 
  }
\email{andreas.triantafyllopoulos@uni-a.de}

\begin{document}

\maketitle

\begin{abstract}
Chronic obstructive pulmonary disease (COPD) causes lung inflammation and airflow blockage leading to a variety of respiratory symptoms; it is also a leading cause of death and affects millions of individuals around the world.
Patients often require treatment and hospitalisation, while no cure is currently available. 
As COPD predominantly affects the respiratory system, speech and non-linguistic vocalisations present a major avenue for measuring the effect of treatment. 
In this work, we present results on a new COPD dataset of 20 patients, showing that, by employing personalisation through speaker-level feature normalisation, we can distinguish between pre- and post-treatment speech with an unweighted average recall (UAR) of up to 82\,\% in (nested) leave-one-speaker-out cross-validation.
We further identify the most important features and link them to pathological voice properties, thus enabling an auditory interpretation of treatment effects.
Monitoring tools based on such approaches may help objectivise the clinical status of COPD patients and facilitate personalised treatment plans.
\end{abstract}

\glsresetall

\section{Introduction and related work}
\label{sec:intro}

\toappear{INTERSPEECH 2022}

Chronic obstructive pulmonary disease (COPD) is a respiratory disease characterised by a chronically inflamed and obstructed airway. The long-term exposure to damaging irritants, especially cigarette smoke, is the predominant 
cause of COPD~\citep{Yoshida07-POC} -- the third leading cause of death after ischemic heart disease and stroke~\citep{WHO20}. Its prevalence in Europe ranges between 15 and 20\,\% for adults aged over 40 years~\citep{Rycroft12-EOC}. Thus, the socio-economic burden of COPD is immense and results in a high consumption of clinical resources and overall costs to society~\citep{Guarascio13-TCA}. 
While the clinical appearance of COPD varies, the main symptoms are chronic and progressive dyspnea, as well as coughing. Sometimes, patients suffer from a severe deterioration of their respiratory symptoms, called exacerbation, which leads to a need of intensified therapy, often under in-patient or even intensive care treatment.
Besides treatment and prevention of exacerbation, the main objective of COPD therapy is to provide symptomatic relief. For this purpose, patients receive (mainly inhalative) medication, such as inhaled corticosteroids (ICS), long-acting muscarinic antagonists (LAMA) or long-acting beta-2-agonists (LABA) -- often administered in combination. Sometimes, further systemic medication, such as systemic corticosteroids, is needed. 
The intensity of treatment depends on the severity of COPD, which is classified according to the recommendations of the Global Initiative of Chronic Obstructive Lung Disease. 
This classification is based on the severity of airflow limitations and takes into account symptoms and risks of exacerbation~\citep{Singh19-GSF}.
Spirometric examinations and clinical findings are still the relevant diagnostic tools for COPD. Especially for acute exacerbation, clinical practitioners must rely on clinical examination to monitor treatment success or failure. 
So far, there are no technical tools to objectivise clinical symptoms of COPD, especially during an exacerbation episode.
In order to accelerate diagnosis and distinguish different respiratory illnesses, new artificial intelligence (AI)-assisted monitoring tools can be helpful.
For instance, new data shows the potential use of AI speech analysis for respiratory diseases, such as COVID-19~\citep{Schuller20-CAC}. 
It is intuitive to expand these findings to COPD, since this obstructive airway disease inevitably affects speech and non-linguistic vocalisations, especially during exacerbation.

Due to the widespread prevalence of COPD and its negative effect on public health, voice-based digital detection and monitoring tools have recently attracted increased attention~\citep{Mohamed14-VCI, Crooks17-CCM, Nathan19-AOC, Ashraf20-VSA, Merker20-DEA, Cleres21-LAV, Farrus21-SSS}.
These utilise different types of vocalisations, such as breathing~\citep{Ashraf20-VSA}, coughing~\citep{Crooks17-CCM}, sustained vowels~\citep{Merker20-DEA}, or read/free speech~\citep{Mohamed14-VCI, Cleres21-LAV, Merker20-DEA} to distinguish between COPD patients and healthy individuals and different states of COPD (such as `stable' vs exacerbation \citep{Merker20-DEA}).
Yet, most of these studies are merely identifying acoustic descriptors that are correlated with COPD
and do not build an automatic detection tool.
%
Instead,
we focus on developing a machine learning (ML)-based voice evaluation tool that distinguishes between pre- and post-treatment states of patients after exacerbation, based on read speech.
We investigate a set of different feature sets, segmentation strategies, and normalisation procedures.
It turns out  that speaker-level feature normalisation is crucial for obtaining good performance -- demonstrating that personalisation is key for this application.
This paves the way for more advanced personalisation techniques which adapt to specific speakers using either speaker-dependent models~\citep{rudovic2018personalized} or test-time adaptation~\citep{triantafyllopoulos2021deep}.
Furthermore, we try to interpret the most important features, which helps characterise the effect of exacerbation and the corresponding benefits of treatment on the speakers' voice.

The remainder of our contribution is organised as follows:
\cref{sec:data} describes the dataset used in this study.
\cref{sec:experiments} outlines our experimental protocol, followed by our results and  accompanying discussion and interpretation  in \cref{sec:results}.
The work ends with some concluding remarks in \cref{sec:conclusion}.

\section{Dataset}
\label{sec:data}

Our COPD dataset was recorded at the University Hospital Augsburg between October 2020 and December 2021. 
Patients were recruited soon after hospitalisation, if possible, already in the emergency department or the intensive care unit. 
Only patients older than 18 years, able to sit and read a short text without any need of ventilation during the time of recording, were included, resulting in: 
20 (11 male / 9 female) subjects, aged 48 to 82 (median: 70), with a median of 37.5 pack years\footnote{Pack years are calculated by multiplying the number of packs of cigarettes smoked per day by the number of years the person has smoked.} and an
improvement (post- to pre-treatment) between 0 and 7 (median: $2.39$) on a modified  Borg scale, a subjective assessment of dyspnea from 0 (worst) to 10 (best).


The local ethics committee approved the study on June 24, 2020 (BKF 2020-34).
Two recordings -- one pre-treatment, one post-treatment -- with a distance of 1 to 16 days (median: 5) took place bedside with the  portable recorder H5 from Zoom\textsuperscript{\textregistered}
and a Sony lapel mic ECM-144.
All patients obtained their standard and individual home medication, mainly 
LABA and LAMA, sometimes in addition to ICS and systemic corticosteroids. 
Some patients needed non-invasive ventilation, which was paused for the time of recording.



Patients  were required to produce: 
\begin{enumerate*}[label=\alph*)]
    \item a set of sustained vowels (\textit{/a:/}, \textit{/e:/}, \textit{/i:/}, \textit{/o:/}, \textit{/u:/});
    \item a few  spontaneous utterances;
    \item (forced) coughing;
    \item breathing; 
    \item  reading 
    \textit{Der Nordwind und die Sonne} [\textit{The Northwind and the Sun}] (NuS).
\end{enumerate*}
We assume no or a negligible habituation effect, as the second recording took place days after the first one.
Here, we deal only with NuS:
it is longer than other sound types (median: 53\,s), thus, resulting in more samples after segmentation, which is beneficial for ML algorithms.
This was confirmed by preliminary experiments with the other sounds.


\section{Experimental setup}
\label{sec:experiments}

\textbf{\textsc{Segmentation:}} 
\ac{NuS} is too long as a unit of analysis for most speech processing applications, which typically operate on shorter segments.
For a segmentation yielding shorter units, we use the \ac{MAUS} system~\citep{schiel1999,kisler2017multilingual}.
It  utilises forced alignment to derive word and phone boundaries from the text transcriptions, which in the case of read speech is trivially available.
We experiment with two different types of segmentation resulting in two different units of analysis:
\textbf{Word units}, where we keep the original word boundaries returned by \ac{MAUS}, resulting in a total of $18\times2\times188=3888$ segments; and 
\textbf{Phrase units}, where we segmented the story into $20$ prosodic phrases for a total of $18\times2\times20=720$ segments.

\noindent
\textbf{\textsc{Features:}} 
We extract a set of acoustic descriptors per unit (for both words and phrases) that can be used to distinguish between pre- and post-treatment COPD speech,
employing both expert, handcrafted features,   and learnt representations of deep neural networks, thus contrasting the two dominant ongoing trends in speech processing applications:

{\textbf{\acs{egemaps}}: The \ac{egemaps}~\citep{eyben2015geneva} is a small set of (88) interpretable acoustic parameters that has previously been shown to contain relevant information for respiratory diseases, such as COVID-19~\citep{bartl2021voice}.
\ac{egemaps} is extracted using openSMILE~\citep{eyben2010opensmile}.}

{\textbf{\acs{compare}}: The \ac{compare} is a large-scale feature set (6373) that has been successfully used for several computational paralinguistics tasks, beginning with the 2013 Interspeech Computational Paralinguistics Challenge~\citep{schuller2013interspeech}, also extracted using openSMILE~\citep{eyben2010opensmile}.}


{\textbf{\wav}: 
Substantial progress has been seen through the use of models trained on vast amounts of data with self-supervised methods.
We use a variant of \textsc{wav2vec2.0}~\citep{baevski2020wav2vec}, pre-trained on 53 languages -- including German~\citep{conneau21crosslingual}.
The model operates on raw audio and returns contextualised representations roughly corresponding to 25\,ms of audio with a stride of 20\,ms, which we subsequently average over the time dimension to obtain the final 1024-dimensional embeddings.
}


\noindent
\textbf{\textsc{Normalisation:}} We experiment with three different normalisation procedures.
In all cases,  z-score normalisation on each feature is performed separately; what changes is the set over which we compute and apply statistics.

\textbf{Global}: As a standard baseline, we normalise the data on a global basis -- for each fold in our cross-validation setup, we compute feature statistics on the training set, and subsequently use those to normalise the development and testing partitions.
This is the prevailing type of normalisation.

\textbf{Word/Phrase-level}: We experiment with a normalisation procedure targeted at the respective \emph{unit of analysis}.
Using read speech recordings, we collect identical audio content for each speaker.
This content, however, is influenced by non COPD-related factors, which can be abstracted away by normalising each word or phrase unit independently (using data from all speakers).

\textbf{Speaker-level}: Given our expectation that there are individual differences in the manifestation of COPD in human vocalisations, we employ a speaker-level normalisation procedure in an attempt to abstract away from them.
This is based on computing (and applying) mean and standard deviation normalisation \emph{independently} for each speaker, that is, using all their data to compute parameters irrespective of whether they are part of the training, development, or test partition.
As such, this form of normalisation assumes oracle knowledge of the identity of each speaker.
We consider this a realistic assumption for \textit{personalised} digital health applications in controlled conditions.

\noindent
\textbf{\textsc{Classifier:}}  We use \acp{SVM} where we optimise the cost parameter (\{$.0001$, $.0005$, $.001$, $.005$, $.01$, $.05$, $.1$, $.5$, $1$\}) and kernel function (\{linear, polynomial, \ac{RBF}\}) in a grid search manner.
These parameters are always optimised on the development partition.

\noindent
\textbf{\textsc{Evaluation protocol:}} As the size of our dataset is limited (20  speakers), we use leave-one-speaker-out cross-validation, whereby data from every speaker is used exactly once for testing, each time using the data of all other speakers for training.
For each fold, we perform \emph{nested cross-validation} for optimising \ac{SVM} parameters by further splitting the training speakers into two speaker-disjoint sets.
Once the optimal set of parameters has been identified (based on development set performance), we train a final model on the entire training data for each fold.

\noindent
\textbf{\textsc{Metrics}:} We use unweighted average recall (UAR), the mean of the diagonal cells in the confusion matrix in percent.
This balances the sensitivity and specificity of both classes, which in our case are both equally important (\ie we have no `positive' and `negative' class)\footnote{As we always have the same number of instances for both classes, UAR is identical with accuracy, as well as with (sensitivity + specificity) / 2, and differs from \fscore throughout just by $\pm1$ percent.}.
We differentiate between three different ways of computing UAR: unit-, story-, and speaker-level.
The first (\textbf{U\textsubscript{UAR}}) quantifies how well the model works over individual \textbf{units}  (instances); 
the second (\textbf{ST\textsubscript{UAR}}), how well it classifies speakers into pre- and post-treatment states after aggregating all individual predictions for each \textbf{story} (the major focus of our work); the third (\textbf{SP\textsubscript{UAR}}), how well the model works for an individual \textbf{speaker} by computing performance over only their instances.
Given a set of speakers $\{s_{1},...,s_{S}\}$ (with $S=20$ being the total number of speakers), each producing  the NuS story twice (corresponding to the two classes \{(b)efore, (a)fter\}), with each story segmented into $N_{u}$ units (\ie words/phrases) resulting in a total of $N$ units (3888/720) overall and $N_s$ units per speaker (216/40), we generate a total of $N$ unit-level predictions $\hat{y}_i$ using the setup outlined above.
We define the different evaluation protocols as follows:
\begingroup
\setlength{\abovedisplayskip}{1pt}
\setlength{\belowdisplayskip}{0pt}
\setlength{\abovedisplayshortskip}{0pt}
\setlength{\belowdisplayshortskip}{0pt}
\begin{align}
    & \text{U}_{\text{UAR}} = \frac{1}{2} \sum_{c \in \{a, b\}} \frac{|i \in [N]: y_{i}=c, \hat{y}_{i}=c|}{|i \in [N]: y_{i}=c|}\nonumber\\
    & \text{ST}_{\text{UAR}} = \frac{1}{2} \sum_{c \in \{a, b\}} \frac{|i \in [S]: y_{i}=c, \underset{j \in [N_s]}{\text{maxvote}} (\hat{y}_{j})=c|}{|i \in [S]: y_{i}=c|}\nonumber\\
    & \text{SP}_{\text{UAR}} = \frac{1}{2} \sum_{c \in \{a, b\}} \frac{|i \in [N_{s}]: y_{i}=c, \hat{y}_{i}=c|}{|i \in [N_{s}]: y_{i}=c|}\nonumber
\end{align}
\endgroup

\section{Results and discussion}
\label{sec:results}
\begingroup
\begin{table}[t]
    \caption{
    U\textsubscript{UAR} and ST\textsubscript{UAR} using leave-one-speaker-out cross-validation with 95\,\% CIs.
    }
    \label{table:NuS}
    \centering
    \setlength{\tabcolsep}{2pt}
    \resizebox{\columnwidth}{!}{
    \begin{tabular}{cc|cccc}
    \toprule
    & & \multicolumn{2}{c}{\textbf{Word units}} & \multicolumn{2}{c}{\textbf{Phrase units}}\\
    \textbf{Normalisation} & \textbf{Features} & \textbf{U\textsubscript{UAR}[\%]} & \textbf{ST\textsubscript{UAR}[\%]} & \textbf{U\textsubscript{UAR}[\%]} & \textbf{ST\textsubscript{UAR}[\%]} \\
    \midrule
    & \acs{egemaps} & 56 (54-57) & 60 (45-75) & 56 (52-59) & 52 (37-68)\\
    Global &\acs{compare} & 52 (51-54) & 60 (44-76) & 53 (49-56) & 50 (34-66)\\
    & \wav & 54 (52-55) & 45 (29-61) & 59 (55-62) & 65 (50-80)\\
    \midrule
    & \acs{egemaps} & 56 (55-58) & 60 (46-75) & 59 (56-62) & 57 (42-73)\\
    Word/Phrase & \acs{compare} & 53 (52-55) & 62 (47-78) & 53 (49-57) & 55 (39-70)\\
    & \wav & 54 (52-55) & 50 (34-65) & 55 (51-59) & 57 (42-72)\\
    \midrule
    & \acs{egemaps} & 60 (58-62) & 68 (53-82) & 63 (60-66) & \colorbox{lightgray}{\textbf{80 (67-92)}}\\
    Speaker & \acs{compare} & 53 (52-54) & 62 (48-79) & 55 (51-58) & 55 (40-70)\\
    & \wav & 59 (58-61) & \colorbox{lightgray}{\textbf{82 (70-94)}} & 66 (63-70) & 78 (66-90)\\
    \bottomrule
    \end{tabular}
    }
\end{table}
\endgroup

\begin{figure*}[t]
    \centering
    \includegraphics[width=.99\textwidth]{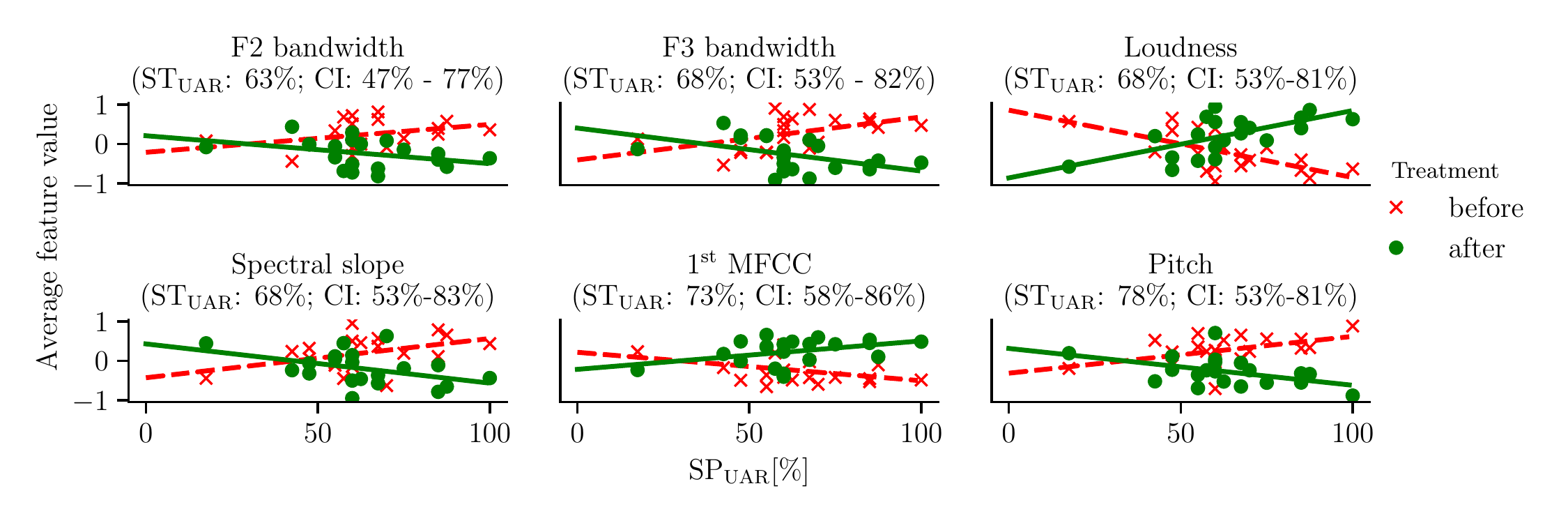}
    \caption{
    Average normalised feature values (taken over all utterances of a speaker) vs SP\textsubscript{UAR}
    for 6 out of 10 most important features as computed by SHAP values;
    %
    values before (red, dashed, cross) and after (green, continuous, point) treatment; 
    identical position per speaker on the x-axis across all plots, with crosses and points sharing the same x-coordinate corresponding to the same speaker; 
    thus, the rightmost points on the x-axis show the speaker normalised mean feature values corresponding to the best performing speaker with 98\,\% (39/40) of their phrases classified correctly
    with \acs{egemaps} and speaker-level normalisation.
    Subtitles show ST\textsubscript{UAR} (total and 95\,\% CI) when using single  features for classification (see text for discussion).
    Linear regression lines 
    fitted to better highlight trends.
    }
    \label{fig:features}
    \vspace{-0.5cm}
\end{figure*}

Overall results are presented in \cref{table:NuS} with U\textsubscript{UAR} and ST\textsubscript{UAR} computed over all instances and corresponding 95\,\% CIs computed over 1000 bootstrap samples; ST\textsubscript{UAR} shows a bigger range over U\textsubscript{UAR}, because it is computed with far less samples (36 vs 3888/720).
Story-level performance is highest for {\wav} and word-level segmentation with a UAR of 82\,\% (CI: 70\,\%-94\,\%) followed by \ac{egemaps} and phrase-level segmentation (ST\textsubscript{UAR}: 80\,\%; CI: 67\,\%-92\,\%) -- both using speaker-level normalisation, with phrase-level \wav ~features trailing close behind (ST\textsubscript{UAR}: 78\,\%; CI: 66\,\%-90\,\%).
The best result without subject-level normalisation is obtained with \ac{compare} with word-level segmentation and normalisation with a ST\textsubscript{UAR} of 62\,\% (CI: 47\,\%-78\,\%) -- a large drop over subject-level normalisation which showcases the need for personalisation.

%
%




\emph{Interpretability} is a necessary requirement for digital health applications in order to explain the outcomes to patients and medical practitioners.
As \ac{egemaps} yields near-top performance, with the features also being easily interpretable due to their expert-designed nature, we focus our subsequent analysis on this setting.
The 95\,\% CI for males (52\,\%-75\,\%) showed substantial overlap to that of females (53\,\%-72\,\%), indicating that the classifier performs approximately equal for both genders. 
We further analysed the performance \wrt the available subject metadata.
We first divided speakers into two subsets: those whose individual performance exceeds the U\textsubscript{UAR} performance of 63\,\% (the phrase-level U\textsubscript{UAR} for \ac{egemaps}  when using instances from all speakers, see  \cref{table:NuS}), and those whose performance falls below that threshold, and subsequently compared the 95\,\% CIs of the different metadata for those two speaker groups.
This comparison revealed that models work better for subjects which have a higher post-to-pre-treatment difference in the BORG scale ([2.12-4.75] vs [1.92-3.16]), are of a higher age ([63-73] vs [58-67] years), and have smoked more pack years ([39-75] vs [28-44]).
All these factors might contribute to a worse clinical condition with subjects subsequently gaining more from treatment, thus, accordingly leading to bigger changes in their voice characteristics and making it easier to distinguish between their pre- and post-treatment states.

We further analyse the features that have the largest impact on classifier decisions in order to characterise the impact of treatment on patients' voices.
To that end, SHAP (\underline{SH}apley \underline{A}dditive ex\underline{P}lanations)~\citep{Lundberg17-AUA}
has emerged as a powerful tool for extracting feature importance values for individual predictions.
SHAP is based on Shapley values, whose theoretical definition for each feature relies on building surrogate models on all potential feature subsets, and taking the expectation of model output differences for all subsets including the target features vs the same subsets but excluding that feature \citep{Lundberg17-AUA}.
These values can then be aggregated over an entire (test) dataset to derive global feature importance values that can be used to interpret model behaviour.
We focus on the ten most important features, defined  by their average SHAP values. 
We computed the mean of each feature separately (normalised using subject-level normalisation) for each subject and phrase for pre- and post-treatment.
We then plotted the resulting $18\times2=36$ points relative to the SP\textsubscript{UAR} corresponding to each speaker.
This allows us to compare how these individual features change after treatment, but also to relate this change to speakers for which the prediction fails.
To provide a better understanding of the effectiveness of each individual feature, we additionally used each of them in isolation to train a (new) model with the same experimental setup discussed before.
Note that this might result in a different `measure' of feature performance as features behave differently in isolation vs in the presence of other (potentially correlated) features~\citep{Batliner20-PIA}.
In \cref{fig:features}, we show the ST\textsubscript{UAR} (and 95\,\% CIs) obtained for each of them.

Due to space limitations, we only include 6 of those:
As four of the original ten were merely functionals of pitch (mean, median, 20\textsuperscript{th} and 80\textsuperscript{th} percentiles -- measured in semitones), and all of them showed a similar trend, we only provide the best performing one, the mean.
Moreover, we exclude the worst-performing feature, the \textbf{3\textsuperscript{rd} MFCC} (ST\textsubscript{UAR}: 60\,\%),  
showing slightly lower values for low performing subjects.\footnote{
We know from other types of atypical speech 
\citep{Hoenig14-AMO}
that the vowel space is centralised, due to  a less tense and less precise articulation. This might be the case here as well.}
\cref{fig:features} shows the remaining six features.
The \textbf{bandwidth of the 2\textsuperscript{nd} formant F2} (ST\textsubscript{UAR}: 63\,\%) and the \textbf{bandwidth  of the 3\textsuperscript{rd} formant F3} (ST\textsubscript{UAR},  68\,\%), above left and second left in \cref{fig:features}, are computed from the roots of the Linear Predictor (LP) coefficient polynomial.
Dysphonic speakers display a broader formant bandwidth \citep{Ishikawa20-BAA} 
meaning  higher formant dispersion and mutual masking of neighbouring formants and by that, vowels \citep{Cheveigne99-FBA}.

\textbf{Loudness} (ST\textsubscript{UAR}: 68\,\%, above right) is partially controlled by transglottal airflow~\citep{baker2001control}, which is constricted by COPD~\citep{Singh19-GSF}; thus, patients before treatment produce on average utterances of lower loudness; this can be seen when we compare the declining line for before with the rising line for after in \cref{fig:features}.\footnote{ 
Note that we used a lapel microphone that prevents varying distances within the same recording session; yet, there might be slight differences across sessions which can constitute an intervening factor that cannot be fully controlled.}  

The \textbf{spectral slope} (ST\textsubscript{UAR}: 68\,\%), below left in \cref{fig:features},  is computed by fitting an OLS estimator to the logarithmic power spectrum and is thus steeper when the higher frequencies have more energy than the lower ones.
This is opposite to the \textbf{1\textsuperscript{st} MFCC} (ST\textsubscript{UAR}: 73\,\%),   below middle, that can be interpreted as the inverse spectral slope, as it is a weighted ratio of the lower to the higher frequencies.
These features  show opposing trends, with the 1\textsuperscript{st} MFCC increasing and the slope decreasing after treatment -- indicating that the ratio of higher to lower frequencies decreases after treatment.
A higher ratio of  higher to lower frequencies  can be interpreted as higher  breathiness~\citep{Hillenbrand96-ACO}, which is then reduced through treatment.
Note that higher breathiness might go together with decreased loudness \citep{Soedersten90-GCA}.

%

\textbf{Pitch} (measured in semitones,  ST\textsubscript{UAR}: 78\,\%) is the most effective \textit{power feature} \citep{Batliner20-PIA}. 
It shows a strong lowering trend after treatment -- in contrast to \citet{Merker20-DEA}, who found that F0 increases for subjects with stable COPD compared to those with exacerbation (mean: 190\,Hz vs 154\,Hz).
A potential confounder is that irregular phonation caused by exacerbated COPD can lead to more errors in pitch estimation 
by missing some voiced segments, especially in laryngealised (creaky) parts \citep{Batliner93-MAC,Keating15-APO}.
By default, openSMILE excludes segments with a value of 0 in its calculation of the mean; thus, the lower pitch values post-treatment compared to pre-treatment (mean: 150\,Hz vs 160\,Hz) might be due to less irregular phonation.
When including 0-valued segments in the calculation of the mean, we obtain higher values of F0 post-treatment (mean: 89\,Hz vs 85\,Hz), similar to \citet{Merker20-DEA}.
We further investigated this hypothesis by comparing the average voiced segments per second before (mean: 24\,\%) and after treatment (mean: 28\,\%); indeed the proportion of segments detected as voiced increases.
As the read text is identical  in both conditions, (unaccounted) pitch estimation errors remain a plausible explanation for the difference between our findings and those of \citet{Merker20-DEA}.
Yet, there might be a `cocktail' of intervening factors: especially before treatment, patients are  more unsettled and stressed, thus both speech pathology and psychological state might result in strained  voice and higher pitch, and at the same time, in irregular voice partly (mis-) recognised as unvoiced.
In the post-stage, speech pathology is weakened and at the same time, the patients  are relieved and more relaxed, and all this might result in a less strained voice and lower pitch.

Summing up our interpretation:
Subjects after treatment show improved articulatory precision (as shown by the
3\textsuperscript{rd} MFCC and F2/F3 bandwidth),  decreased airflow blockage (as shown by the increase in loudness), decreased breathiness (as shown by the spectral slope and 1\textsuperscript{st} MFCC), and more regular phonation (thus less pitch errors) and, by that, lowered `regular' pitch  -- findings which are consistent with the expected decrease in symptomatology;
as for a comparable voice quality spectrum for Parkinson's disease, see  \citep{Cernak17-COV}.
Naturally, although our present interpretation is consistent with  previous phonetic research and  with medical expectations, it should be evaluated on a larger sample size and tested against human evaluations.

\section{Conclusion}
\label{sec:conclusion}

We demonstrated that read passages can be successfully utilised to distinguish between pre- and post-treatment states of COPD patients. Using a variety of handcrafted and learnt features, we were able to achieve a top UAR of 82\,\% (95\,\% CI: 70\,\%-94\,\%) in a leave-one-speaker-out setup. Speaker-level normalisation proved to be  crucial, as it removes speaker-related effects, which prevents  generalisation; without it, performance reached a maximum of 62\,\%. Future work could be directed to more data efficient personalisation techniques, which do not require patient data to normalise with, as well as to detecting COPD in the presence of other respiratory diseases, such as COVID-19.


\section{Acknowledgements}
This work has received funding from the DFG's Reinhart Koselleck project No.\ 442218748 (AUDI0NOMOUS) and from the EU’s Horizon 2020 grant agreement No.\ 826506 (sustAGE).

\newpage
\section{\refname}
 \printbibliography[heading=none]

\end{document}